\documentclass[lettersize,journal]{IEEEtran}
\usepackage{ifpdf}
\usepackage{booktabs}

\usepackage{cite}

\ifCLASSINFOpdf
\usepackage[pdftex]{graphicx}
\else
\usepackage[dvips]{graphicx}
\fi
\usepackage{amsmath}
\usepackage{algorithmic}

\usepackage{array}

\ifCLASSOPTIONcompsoc
\usepackage[caption=false,font=normalsize,labelfont=sf,textfont=sf]{subfig}
\else
\usepackage[caption=false,font=footnotesize]{subfig}
\fi
\usepackage{fixltx2e}

\usepackage{stfloats}
	\ifCLASSOPTIONcaptionsoff
	\usepackage[nomarkers]{endfloat}
	\let\MYoriglatexcaption\caption
	\renewcommand{\caption}[2][\relax]{\MYoriglatexcaption[#2]{#2}}
	\fi
	\usepackage{url}

	\usepackage{bm}
	\usepackage{enumitem}
	\usepackage{color}
	\usepackage{soul} 
	\usepackage{color, xcolor} 
	\renewcommand{\figurename}{Fig. }
	\usepackage{makecell}
	\usepackage{multirow}

\begin{document}
		%
		\title{ Transient Synchronization Stability Analysis of SG-DFIG Parallel System Considering \\  Complete LVRT Processes}
		
		%
		%
		\author{Hongsheng~Xu,
			Meng~Zhan,~\IEEEmembership{Senior Member,~IEEE},
			Jia Hu,
			Huiyu Shang,
			Qingfeng Yue
			\thanks{
				This work was supported by Smart Grid-National Science and Technology Major Project (2025ZD0807200). {\it{(Corresponding author: Meng Zhan.)}}
				
				Hongsheng Xu and Meng Zhan are with the State Key Laboratory of Advanced Electromagnetic Engineering and Technology, Hubei Electric Power Security and High Efficiency Key Laboratory, School of Electrical and Electronic Engineering, Huazhong University of Science and Technology, Wuhan 430074, China (e-mail: xhs@hust.edu.cn;~zhanmeng@hust.edu.cn).
				
				Jia Hu, Huiyu Shang and Qingfeng Yue are with the Guangzhou Power Supply Bureau of Guangdong Power Grid Co,. Ltd, Guangzhou, 510620, China (e-mail: 15623511398@163.com; 51396593@qq.com; 497604092@qq.com)

			}
		}

	\maketitle
	\begin{abstract}
Although a large amount of work has been devoted to detailed electromagnetic transient simulation in analyzing transient synchronization stability (TSS) of hybrid systems containing renewable energy equipment and synchronous generator (SG), the underlying mechanism considering complete low-voltage ride-through (LVRT) processes of renewable energy equipment remains to be studied. Taking the SG and doubly fed induction generator (SG-DFIG) parallel system as an objective, this work divides its transient processes into four different stages: pre-fault (stage 1), during-fault (stage 2), early post-fault (stage 3), and late post-fault (stage 4), based on the LVRT of the DFIG, and establishes a transient model to describe the complete 4-stage LVRT processes. By studying the condition for entering the LVRT, it is found that vast majority of faults can cause the DFIG to enter the LVRT and make the parallel system exhibit the sequential switching characteristics. Similar to the SG-SG parallel system, which can be reduced to a single SG and described by a second-order swing equation, a unified generalized swing equation (GSE) under different parameters for different stages 1, 2, and 3 is derived. Therefore, the transient stability of the parallel system can be dealt with easily, and further, an improved equal area criterion method considering two additional effects of frequency jump and nonlinear damping is proposed to evaluate the TSS. These GSE-based theoretical analysis results are all supported by extensive hardware-in-the-loop experiments and simulations. Obviously, this work provides a clearer physical picture for the TSS mechanism of the hybrid system considering complete LVRT processes, and makes a closer connection with the transient stability of traditional power systems dominated by SG.
	\end{abstract}


	%
	\IEEEpeerreviewmaketitle

	\section*{Nomenclature}
	\begin{itemize}[leftmargin = 63pt]
		\item[$\vec{U}_{\rm{\mathit{wt}}}$, $\vec{E}_{\rm{\mathit{sg}}}$] Terminal voltage of DFIG and internal voltage of SG.
		\item[$\vec{I}_{\rm{\mathit{s}}}$, $\vec{I}_{\rm{\mathit{r}}}$, $\vec{I}_{\rm{\mathit{wt}}}$] Stator current, rotor current and DFIG total output current.		
		\item[${U}_{\rm{\mathit{wt}}}$,${E}_{\rm{\mathit{sg}}}$] Amplitude of terminal voltage of DFIG and amplitude of internal voltage of SG.
		\item[$i_{\rm\mathit{wtd}}$, $i_{\rm\mathit{wtq}}$] $dq$ axis components of output current of DFIG.

		\item[$u_{\rm \mathit{wtd}}$, $u_{\rm\mathit{wtq}}$] $dq$ axis components of terminal voltage of DFIG.
		\item[$\theta _{\rm \mathit{pll}}$,$\theta _{\rm \mathit{sg}}$ ] Output angles of PLL and SG in three-phase stationary $abc$ reference frame, respectively.
		\item[$\varphi _{\rm \mathit{pg}}$,$\omega _{\rm \mathit{pg}}$ ] Angle and frequency differences between PLL and SG, respectively.
		
		\item[$\varphi_I$] Angle difference between $\vec{I}_{r}$ and the $d_{pll}$-axis of PLL.
		
	    \item[${\varphi _{cr}}$] Critical clearing angle.

		\item[$k_{\rm \mathit{pw}}$, $k_{\rm \mathit{iw}}$]  Proportional and integral (PI) parameters of RSC.
		\item[$k_{\rm \mathit{ppll}}$, $k_{\rm \mathit{ipll}}$]  PI parameters of PLL.
		\item[$K_{\rm \mathit{e}}$, $K_{\rm \mathit{ramp}}$]  Reactive current ratio coefficient and ramp rate.

		\item[${P}_{\rm{\mathit{wt}}}$,  ${P}_{\rm{\mathit{sg}}}$]  Output active powers of DFIG and SG, respectively.

		\item[$\omega _{\rm \mathit{pll}}$, $\omega _{\rm \mathit{sg}}$, $\omega _{\rm \mathit{r}}$] Frequencies of PLL, SG, and DFIG rotor speed, respectively.
		
		\item[$X_\mathit{m}$, $X_\mathit{s}$] Mutual reactance and stator reactance.
		\item[$Z_\mathit{wt}$, $Z_\mathit{sg}$] Line reactances of DFIG and SG, respectively.
		\item[$Z_\mathit{L}$] Constant impedance load.
		
		\item[$A,B,C,D$] Amplitude coefficients defined in (\ref{eq_coe}).
		\item[$\phi _a,\phi _b,\phi _c,\phi _d$] Phase coefficients defined in (\ref{eq_coe}).

		\item[${P_{\rm{\mathit{Meq}}}}$,${P_{\rm{\mathit{Teq}}}}$,${D_{\rm{\mathit{eq}}}}$] Equivalent mechanical power, electromagnetic power, and damping, respectively.
		
		\item[$K_{eq},\gamma,m,n$] Constant coefficients defined in (\ref{eq_coePteq}).
		
		\item[$t_\mathit{f}$, $t_\mathit{c}$, $t_\mathit{r}$ ] Times for the fault occurrence, clearing, and  ramp-ending, respectively.
		
		\item[$1,2,3,4$] Subscripts of stages 1, 2, 3, and 4, for pre-fault, during-fault, early post-fault, and late post-fault, respectively.
		
	\end{itemize}

	\section{Introduction}
	%
	%
	%
	%

	\IEEEPARstart{W}{ith} increasing proportion of renewable energy, hybrid systems dominated by synchronous generators (SG) and renewable energy power generation (REPG) equipment have gradually become the main form \cite{Ref_2,Ref_5}. Its transient stability is one of central problems in the renewable energy dominated  power system. As a traditional power generation equipment, the SG has strong overcurrent capacity, and in the transient stability study it is usually described by the second-order swing equation (SE) \cite{Ref_kundur}. However, due to insufficient overcurrent capacity, the REPG equipment has to adopt transient switching control during fault processes, such as the low-voltage ride through (LVRT). Namely, it should change its internal controllers or parameters to respond the external disturbances properly. The new system shows not only multi-time-scale and strong nonlinearity, but also sequential switching \cite{Ref_HJB}. These distinctive external characteristics make the transient synchronization stability (TSS) analysis of the hybrid system even much more difficult, to be compared with that of traditional power system dominated by the SG \cite{Ref_kundur}.

	In the early stage when the proportion of REPG was low, it is generally believed that the TSS of the hybrid system is dominated by the SG. The REPG equipment is usually regarded as a static component (as an equivalent circuit component \cite{Ref_Tian} or power source \cite{Ref_Shen, Ref_GeX, Ref_ChenL1}), while the rotor dynamics of SG is retained. Based on the traditional analytical methods, such as the equal area criterion (EAC), the impact of REPG equipment on the power angle stability of SG was analyzed. In recent years, the TSS issue of the REPG equipment has gained attention; it is generally believed that (1) non-existence of equilibrium point or (2) transient instability of the phase-locked loop (PLL) on the during-fault stage leads to the out-of-step of REPG equipment, while the following clearing-fault stages are completely ignored \cite{Ref_12,Ref_19,Ref_HeC}. When the PLL  dynamics are ignored, it is found that it introduces a new algebraic constraint to the hybrid system and induces a new form of transient synchronization instability \cite{Ref_HeX}. As the penetration of REPG increases, the transient model of REPG needs to be improved to more accurately reflect its transient process. For some recent works, when the PLL dynamics is taken into account, the phenomenon of multiple equilibrium points was reported \cite{Ref_WangY}. Moreover, as the output active power of REPG increases, it is found that the dominant unstable equipment in the hybrid systems shifts from SG to REPG \cite{Ref_LiX, Ref_WangY2}. This phenomenon has also been observed in the hybrid system considering outer control loop \cite{Ref_ChenL2}. However, most of these existing studies \cite{Ref_hybrid1,Ref_hybrid2,Ref_Qu, Ref_TW,Ref_hybrid3} do not consider the complete LVRT process of the REPG, such as the non-autonomous dynamics during the ramping period, and the impact of each stage in the TSS remains unclear.

	In our recent unified theory on the TSS of the REPG systems \cite{Ref_ZYY2,Ref_xhs,Ref_Shen_Qihao,Ref_han}, we have found that the system considering the complete LVRT process actually shows a high degree of similarity with the SG system. Although the LVRT process involves four stages, including the pre-fault (stage 1), during-fault (stage 2), early post-fault (stage 3), and late post-fault (stage 4), the state at the beginning of stage 3 plays a decisive role in the TSS, based on the quasi-steady-state characteristics of the non-autonomous stage 3.
	Therefore, for the TSS analysis, it actually can be regarded as having only three stages: stages 1, 2, and a virtual stage 3, which just correspond to the pre-fault, during-fault, and clearing-fault stages, respectively, of the traditional power system transient. The dynamical behaviors can be well caught by the same generalized swing equation (GSE) \cite{Ref_16} with different parameters in each stage. In addition, this unified theory fits for diversified REPG equipment of wind and solar energies. Therefore, it is necessary to extend these studies to the hybrid system.

	Taking the doubly-fed induction generator (DFIG) as an example, this paper establishes a switched model including the LVRT dynamics and conducts the TSS analysis and evaluation for the SG-DFIG parallel system. The main contributions of this article are as follows:
	
	1) For the first time, a transient model of the parallel system taking into account the complete 4-stage LVRT process is constructed, and the LVRT conditions are studied. It is found that under most fault conditions, the REPG equipment is prone to the LVRT. 
	
	2) A 3-stage GSE-based model considering stages 1, 2, and 3 is constructed,  based on that the initial moment of stage 3 determines the TSS. For the SG-DFIG parallel system, the phase and frequency mismatches between the SG and DFIG play a crucial role in the analysis, similar to the TSS analysis of the SG-SG parallel system.
	The GSE model still works.
	
	3) In the TSS analysis, considering the \textquotedblleft frequency jump\textquotedblright \hspace{0.2em}and \textquotedblleft nonlinear damping\textquotedblright, an improved EAC method is proposed. This greatly reduces the conservation of the original EAC method. As a result, for the critical clearing angle (CCA), the comparative error is only within -3\%, and for the critical clearing time (CCT), the comparative error is only within -2\%, to be compared with the electromagnetic transient (EMT) simulation results.

	The rest of this article is structured as follows. In Section II, the transient mechanism model of the SG-DFIG parallel system is constructed. In Section III, the condition for entering LVRT is analyzed and the GSE model of the parallel system is derived. In Section IV, the improved EAC-based method is proposed. In Section V, the theoretical and simulation results are widely verified by experiments. Finally, the conclusion is made in Section VI. To be clear, the subscripts 1-4 are used for stages 1-4, for the pre-fault, during-fault, early post-fault, and late post-fault, respectively, in the whole paper.

\section{Transient mechanism model}	
\subsection{Topological Structure and Control} 

		\begin{figure}[!t]
	\centering
	\includegraphics[width=0.7\linewidth]{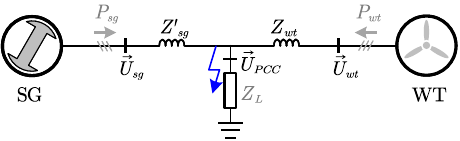}
	\vspace{-0.2cm}
	\caption{Topology diagram of the SG-DFIG parallel system.}
	\label{topology_all}
\end{figure}

	Figure \ref{topology_all} shows the topological structure of the SG-DFIG parallel system, where the SG and the DFIG are connected in parallel to a constant impedance load. This is very common in the two-side-supporting power grid and the power system model simplification, and it has also been widely studied recently \cite{Ref_hybrid1,Ref_hybrid2,Ref_Qu, Ref_TW,Ref_hybrid3}. Here, the infinite bus node does not exist. The SG is regarded as a voltage source, whose dynamics is generally described by the SE \cite{Ref_kundur}. The DFIG is usually regarded as a current source. In the entire fault process, the SG's internal dynamics remains unchanged, while only the network equations and parameters vary. However, for the DFIG under severe faults, e.g., when the terminal voltage is less than 0.8 p.u., it should enter the LVRT according to the grid code \cite{Ref_GB}. To ensure equipment safety and grid-connected operation, in the LVRT the DFIG should adopt sequential switching controls at different stages. Throughout the entire LVRT process, not only do the network equations change, but also the internal dynamics of the DFIG. The parameters are given in Appendix A.

	 To make a direct comparison, \figurename \ref{SG_SG} in Appendix B schematically shows a double-SG parallel system, as a classic topological structure in the traditional power system \cite{Ref_kundur}. It has been widely studied in the TSS analysis and also in the out-of-step splitting etc \cite{Ref_kundur}. It can be reduced to a single SG system described by the SE under the uniform condition. The details can be found in Appendix B.

 Figure \ref{DFIG_control} shows the typical four stages of the DFIG: the pre-fault (stage 1), during-fault (stage 2), early post-fault (stage 3), and late post-fault (stage 4). In stage 1, the DFIG adopts the normal control. The active power control of the machine side converter (MSC) is the rotor speed control (RSC), and the reactive power control selects the constant reactive power injection (i.e., the normal reactive current $i_{rq,norm}$ is a constant). When a severe fault occurs under $U_{wt}$ $<$ 0.8 p.u., the LVRT is triggered, and the DFIG enters stage 2, by adopting the LVRT control to support the terminal voltage. After the fault is cleared, the DFIG enters stage 3 and adopts the Ramp control to restore the active power. When the active power returns to the initial level, the DFIG enters stage 4 and resumes the normal control.

		\begin{figure}[!t]
			\centering
			\includegraphics[width=0.8\linewidth]{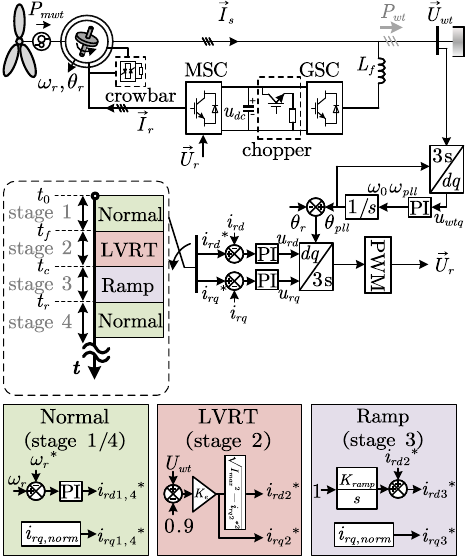}
			\vspace{-0.2cm}
			\caption{Control block diagram of the DFIG considering LVRT processes.}
			\label{DFIG_control}
			\vspace{-0.3cm}
		\end{figure}

To catch the dominant dynamics of this parallel system, we retain the electromechanical dynamics and the PLL synchronization control, under the following assumptions:

1) The SG neglects the excitation dynamics and only considers the second-order rotor dynamics by the SE \cite{Ref_kundur}.

2) The line inductance dynamics is ignored, as the system frequency is around 50 Hz. The frequency operating point is regulated and provided by the inherent damping of the SG, $D_{sg}$ \cite{Ref_kundur}.

3) The flux linkage dynamics of the DFIG is neglected, and the AC current control is ideal (i.e., ${i\rm_{\mathit{rd}}}={i\rm_{\mathit{rd}}}^{*}$ and ${i\rm_{\mathit{rq}}}={i\rm_{\mathit{rq}}}^{*}$) \cite{Ref_xhs}.

4) As the output power of the grid side converter (GSC) accounts for only $20\% \sim 30\%$ of whole power of the DFIG \cite{Ref_Wu}, its impact on the system is relatively small. And as the LVRT and the Ramp controls are executed by the MSC, the GSC is disregarded and only the MSC and the stator branch are considered \cite{Ref_10}. 

\begin{figure}[!t]
	\centering
	\includegraphics[width=0.8\linewidth]{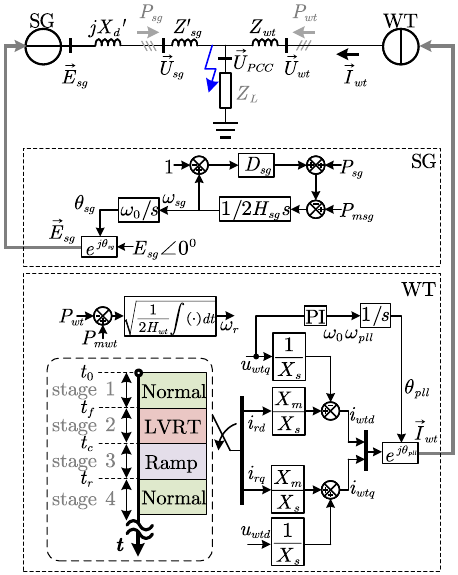}
	\vspace{-0.2cm}
	\caption{Schematic show of mechanism model for SG-DFIG parallel system.}
	\label{mechanism_2Gen}
	\vspace{-0.3cm}
\end{figure}

These assumptions have been carefully examined in the model and simulation comparisons. Therefore, the dominant control block diagram of the DFIG and the SG in the simplified mechanism model are shown in \figurename \ref{mechanism_2Gen}, where the SG is regarded as a voltage source ($\vec{E}_{sg}$), and the DFIG as a current source ($\vec{I}_{wt}$). The details will be explained subsequently.

\subsection{Transient model of stage 1 (and stage 4)}

As the first step, let us study the dynamics of stage 1 (and equivalently, stage 4). 
	Firstly, we derive the network algebraic equations. According to the superposition law of linear circuits, based on the $\vec{E}_{sg}$ and $\vec{I}_{wt}$, the terminal voltage $\vec{U}_{wt}$ of the DFIG and the output current $\vec{I}_{sg}$ of the SG are			
				\begin{equation}
					\label{eq_Cir1}
					\left\{ \begin{array}{l}
	\vec{U}_{wt}=\dfrac{Z_L}{Z_L+Z_{sg}}\vec{E}_{sg}+\left( Z_{wt}+\dfrac{Z_LZ_{sg}}{Z_L+Z_{sg}} \right) \vec{I}_{wt}
	\\
		\vec{I}_{sg}=\dfrac{1}{Z_L+Z_{sg}}\vec{E}_{sg}-\dfrac{Z_L}{Z_L+Z_{sg}}\vec{I}_{wt}\\[3mm]
				\end{array} \right.
				\end{equation}
	where $Z_{sg}$ ($Z_{sg}=Z_{sg}^{\prime} +j{X_d}^{\prime} $) is the impedance from the $\vec{E}_{sg}$ to the point of common coupling (PCC) voltage $\vec{U}_{PCC}$ (here $Z_{sg}^{\prime}$ denotes the impedance from the SG terminal to the PCC and ${X_d}^{\prime}$ denotes the SG $d$-axis transient impedance), and $Z_{wt}$ is the impedance from $\vec{U}_{wt}$ to $\vec{U}_{PCC}$, as shown in \figurename \ref{topology_all}. In addition, without losing generality, $Z_L$ is chosen as a purely resistive load. In this paper, we will mainly study a ground fault at the PCC, and therefore, we set $Z_L$ drop to a very small number in stage 2 and recover at stages 3 and 4.

Above we have obtained the terminal relations between ($\vec{U}_{wt}$, $\vec{I}_{sg}$) and ($\vec{E}_{sg}$, $\vec{I}_{wt}$). Nevertheless, as the DFIG actually controls the rotor current $\vec{I}_{r}$, we have to derive the relations between ($\vec{U}_{wt}$, $\vec{I}_{sg}$) and ($\vec{E}_{sg}$, $\vec{I}_{r}$). When the flux linkage dynamics is ignored, the relation between the rotor current $\vec{I}_{r}$ and the stator current $\vec{I}_{s}$ is
				\begin{equation}
				Z_s\vec{I}_s=Z_m\vec{I}_{r}-\vec{U}_{wt}
	\label{eq_Flux}
				\end{equation}
	where $Z_{s}=jX_{s}$ and $Z_{m}=jX_{m}$, with $X_{\rm\mathit{s}}$ and $X_m$ representing the stator reactance and mutual reactance, respectively.

	Because the output current of the GSC is ignored, the output current $\vec{I}_{wt}$ of the DFIG is equal to the stator current $\vec{I}_{s}$, i.e.,
				\begin{equation}
				\vec{I}_{wt} = \vec{I}_{s}
	\label{eq_current}
				\end{equation}

Combining (\ref{eq_Flux}) and (\ref{eq_current}), the component expressions of the output current $\vec{I}_{wt}$ of the DFIG can be obtained, 
				\begin{equation}
					\label{eq_iwtdq}
					\left\{ \begin{array}{l}
	i_{wtd}=\left( X_mi_{rd}-u_{wtq} \right) /X_s\\
	i_{wtq}=\left( X_mi_{rq}+u_{wtd} \right) /X_s\\
					\end{array} \right.
				\end{equation}
which are illustrated in \figurename \ref{mechanism_2Gen}. 				
				
Next combining (\ref{eq_Cir1})–(\ref{eq_current}), we eliminate the intermediate variable $\vec{I}_{wt}$ and have 
				\begin{equation}
					\label{eq_Cir2}
					\left\{ \begin{array}{l}
	\vec{U}_{wt}=A\angle \phi _a\vec{E}_{sg}+B\angle \phi _b\vec{I}_{r}\\
	\vec{I}_{sg}=C\angle \phi _c\vec{E}_{sg}-D\angle \phi _d\vec{I}_{r}\\
					\end{array} \right.
				\end{equation}
and
\begin{small}
				\begin{equation}
					\label{eq_coe}
					\left\{ \begin{array}{l}
		A\angle \phi _a=\dfrac{Z_LZ_s}{\left( Z_s+Z_{wt} \right) \left( Z_L+Z_{sg} \right) +Z_LZ_{sg}}\\[3mm]
	B\angle \phi _b=\dfrac{\left[ Z_{wt}\left( Z_L+Z_{sg} \right) +Z_LZ_{sg} \right] Z_m}{\left( Z_s+Z_{wt} \right) \left( Z_L+Z_{sg} \right) +Z_LZ_{sg}}\\[3mm]
	C\angle \phi _c=\dfrac{Z_s+Z_{wt}+Z_L}{\left( Z_s+Z_{wt} \right) \left( Z_L+Z_{sg} \right) +Z_LZ_{sg}}\\[3mm]
	D\angle \phi _d=\dfrac{Z_LZ_m}{\left( Z_s+Z_{wt} \right) \left( Z_L+Z_{sg} \right) +Z_LZ_{sg}}\\
					\end{array} \right.
				\end{equation}
\end{small}				
where $A\angle \phi _a$, $B\angle \phi _b$, $C\angle \phi _c$, and $D\angle \phi _d$ are all (constant) complex coefficients depending on the network parameters, and correspondingly $A\sim D$ are their amplitudes, and $\phi _a\sim \phi _d$ are their angles.

It is notable that as the topology of the DFIG differs from that of other REPGs, such as the permanent magnet synchronous generator (PMSG) and photo-voltaic (PV), where the network relation (\ref{eq_Cir1}) is directly applicable, here (\ref{eq_Cir2}) for the DFIG should be used, instead. The only difference between (\ref{eq_Cir1}) and (\ref{eq_Cir2}) comes from their different control variables and the associated coefficients. Therefore, the basic rules and analytical methods in this paper can be applied to the PMSG and PV directly \cite{Ref_ZYY2,Ref_Shen_Qihao,Ref_han}.
 
						\begin{figure}[!t]
	\centering
	\includegraphics[width=0.47\linewidth]{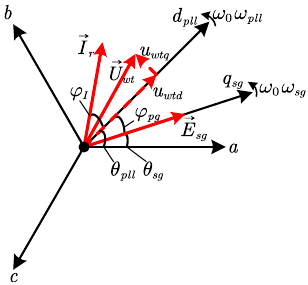}
	\vspace{-0.2cm}
	\caption{Schematic diagram of the $abc$ stationary coordinate frame, the PLL rotating coordinate frame, and the SG rotating coordinate frame.}
	\label{zuobiao}
	\vspace{-0.3cm}
\end{figure}

Figure \ref{zuobiao} shows the major vectors and their phase relations. 
In (\ref{eq_Cir2}), the internal potential $\vec{E}_{sg}$ of the SG and the rotor current $\vec{I}_{r}$ of the DFIG can be expressed as
				\begin{equation}
					\label{eq_ctrolled}
					\left\{ \begin{array}{l}
					\vec{E}_{sg}=E_{sg}\angle \theta _{sg}
\\
\vec{I}_r=\left( i_{rd}+ji_{rq} \right) 1\angle \theta _{pll}=I_r\angle \left( \varphi _I+\theta _{pll} \right) 
					\end{array} \right.
				\end{equation}
where $E_{sg}$ is the constant amplitude of $\vec{E}_{sg}$, $I_r$ ($I_r=\sqrt{{i_{rd}}^2+{i_{rq}}^2}$) represents the amplitude of $\vec{I}_r$,
$\varphi_I$ denotes the angle between the $\vec{I}_{r}$ and the $d_{pll}$-axis of the PLL, and $\theta_{sg}$ and  $\theta_{pll}$ represent the output angle of the SG and the PLL in the three-phase stationary $abc$ reference frame, respectively.

In fact, by using their phase difference $\varphi_{pg}$ and frequency difference $\omega_{pg}$, i.e.,
				\begin{equation}
					\label{eq_wphire}
					\left\{ \begin{array}{l}
\varphi_{pg} = \theta_{pll} -\theta_{sg}
\\
\omega_{pg} = \omega_{pll} - \omega_{sg}
					\end{array} \right.
				\end{equation}
the analysis of this complicated dynamical system can be greatly simplified, as we will see.

Next, based on the coordinate relations in \figurename \ref{zuobiao}, the vector $\vec{U}_{wt}$ in (\ref{eq_Cir2}) can be expressed in the PLL coordinate frame:
				\begin{equation}
					\label{eq_uwtdq}
					\left\{ \begin{array}{l}
	u_{wtd}= AE_{sg}\cos \left( \varphi_{pg}-\phi _a \right) + BI_r\cos \left( \phi _b+\varphi _I \right) \\
	u_{wtq}= -AE_{sg}\sin \left( \varphi_{pg}-\phi _a \right) + BI_r\sin \left( \phi _b+\varphi _I \right)  \\
					\end{array} \right.
				\end{equation}
Therefore, we have the terminal voltage amplitude $U_{wt}$ and the output power $P_{wt}$ of the DFIG,
				\begin{equation}
					\label{eq_PU}
					\left\{ \begin{array}{l}
					U_{wt}=\sqrt{{u_{wtd}}^2+{u_{wtq}}^2}
\\
P_{wt}=u_{wtd}i_{wtd}+u_{wtq}i_{wtq}
					\end{array} \right.
				\end{equation}
Further, according to (\ref{eq_Cir2}), we obtain the output power $P_{sg}$ of the SG as a nonlinear function of $\varphi_{pg}$, i.e.,
				\begin{equation}
					\label{eq_PSG}
					 \begin{aligned}
P_{sg} &= \mathrm{Re}\left[ \vec{E}_{sg}\overline{ \vec{I}_{sg}} \right]
\\
&={CE_{sg}}^2\cos\phi _c -E_{sg}DI_r\cos \left[ \varphi_{pg}+( \phi _d+\varphi _I)\right]
					\end{aligned} 
				\end{equation}
where the horizontal-line symbol above ${ \vec{I}_{sg}}$ represents conjugate, which will be used throughout this paper.

Although the controls of the DFIG vary at different stages, the algebraic relations (\ref{eq_uwtdq}) - (\ref{eq_PSG}) are unchanged. The only difference comes from the rotor current outer loop control and the load resistance, as shown in Figs. \ref{DFIG_control} and \ref{mechanism_2Gen}.

Throughout the entire transient process, the SG always considers the rotor dynamics. Therefore, in stages 1 and 4, the rotor dynamics and the RSC control and PLL in the DFIG are dominant. Based on ${\varphi}_{pg}$ and ${\omega}_{pg}$, we have the following differential algebraic equations (DAEs) of the SG-DFIG parallel system in stages 1 and 4 in (\ref{eq_stage1DE}) and (\ref{eq_stage1AE}):
				\begin{equation}
				\label{eq_stage1DE}
					\left\{ \begin{array}{l}
	\dot{\varphi}_{pg}=\omega _0\omega _{pg}\\
	\dot{\omega}_{pg}=\left( k_{ppll}\dot{u}_{wtq}+k_{ipll}u_{wtq} \right) /\omega _0 -\dot{\omega}_{sg}\\
	\dot{\omega}_{sg}=\left[ P_{msg}-P_{sg}-D_{sg}\left( \omega _{sg}-1 \right) \right] /\left( 2H_{sg} \right)\\
	\dot{\omega}_r=\left( P_{mwt}-P_{wt} \right) /\left( 2H_{wt}\omega _r \right)\\
	\dot{i}_{rd}=k_{pw}\dot{\omega}_r+k_{iw}\left( \omega _r-{\omega _r}^* \right)\\
						\end{array} \right.
				\end{equation}
		where $D_{sg}$ is the damping of the SG, $H_{sg}$ and $H_{wt}$ are the inertia of the SG and DFIG, respectively, $P_{msg}$ and $P_{mwt}$ are the (constant) active power inputs of the SG and DFIG, respectively, $k_{pw}$ and $k_{iw}$ are the proportional and integral coefficients of the RSC, respectively, ${\omega_r}^{*}$ is the reference value of the rotor speed, and $k_{ppll}$ and $k_{ipll}$ are the proportional and integral coefficients of the PLL, respectively;
		\begin{small}
				\begin{equation}
				\label{eq_stage1AE}
					\left\{ \begin{array}{l}
		i_{rq}=i_{rq,norm}\\
	I_r=\sqrt{{i_{rd}}^2+{i_{rq}}^2}\\
	\varphi _I=\arctan \left( i_{rq}/i_{rd} \right)\\
	u_{wtd}= AE_{sg}\cos \left( \varphi_{pg}-\phi _a \right) + BI_r\cos \left( \phi _b+\varphi _I \right)   \\
	u_{wtq}= -AE_{sg}\sin \left( \varphi_{pg}-\phi _a \right) +BI_r\sin \left( \phi _b+\varphi _I \right)  \\
	i_{wtd}=\left( X_mi_{rd}-u_{wtq} \right) /X_s\\
	i_{wtq}=\left( X_mi_{rq}+u_{wtd} \right) /X_s\\
	P_{wt}=u_{wtd}i_{wtd}+u_{wtq}i_{wtq}\\
	P_{sg}={CE_{sg}}^2\cos \phi _c-E_{sg}DI_r\cos \left[ \varphi _{pg}+\left( \phi _d+\varphi _I \right) \right]\\
					\end{array} \right.
				\end{equation}
			\end{small}
where $i_{rq,norm}$ denotes the normal reactive current.

\subsection{Transient model of stage 2} 
In stage 2, the DFIG goes into the LVRT control, the PLL dynamics of the DFIG and the SG rotor dynamics are dominant. Thus, we have the following DAEs  (\ref{eq_stage2DE}) and (\ref{eq_stage2AE}):
				\begin{equation}
				\label{eq_stage2DE}
					\left\{ \begin{array}{l}
	\dot{\varphi}_{pg}=\omega _0\omega _{pg}\\
	\dot{\omega}_{pg}=\left( k_{ppll}\dot{u}_{wtq}+k_{ipll}u_{wtq} \right) /\omega _0 - \dot{\omega}_{sg}\\
	\dot{\omega}_{sg}=\left[ P_{msg}-P_{sg}-D_{sg}\left( \omega _{sg}-1 \right) \right] /\left( 2H_{sg} \right)\\
					\end{array} \right.
				\end{equation}
and			
\begin{small}
				\begin{equation}
				\label{eq_stage2AE}
					\left\{ \begin{array}{l}
		i_{rq}=K_e\left( U_{wt}-0.9 \right) +i_{rq,norm}\\
		i_{rd}\leq \sqrt{{I_{max}}^2-{i_{rq}}^2}\\
		I_r=\sqrt{{i_{rd}}^2+{i_{rq}}^2}\\
		\varphi _I=\arctan \left( i_{rq}/i_{rd} \right)\\
		u_{wtd}= AE_{sg}\cos \left( \varphi_{pg}-\phi _a \right) + BI_r\cos \left( \phi _b+\varphi _I \right) \\
		u_{wtq}= -AE_{sg}\sin \left( \varphi_{pg}-\phi _a \right) + BI_r\sin \left( \phi _b+\varphi _I \right)  \\
		U_{wt}=\sqrt{{u_{wtd}}^2+{u_{wtq}}^2}\\
		P_{sg}={CE_{sg}}^2\cos \phi _c-E_{sg}DI_r\cos \left[ \varphi _{pg}+\left( \phi _d+\varphi _I \right) \right]\\
					\end{array} \right.
				\end{equation}
	\end{small}
		where $K_e$ is the reactive current coefficient and ${I_{max}}$ is the current capacity (usually ${I_{max}}$= 1.2 p.u. is chosen). In the calculation of $i_{rq}$, as the value of $U_{wt}$ in the whole stage 2 is nearly unchanged, in engineering it is usually fixed as the value of $U_{wt}$ at the beginning of stage 2 and thus $i_{rq}$ is set as a constant.

\subsection{Transient model of stage 3} 	
In stage 3, the Ramp control and the PLL dynamics in the DFIG dominate, and the rotor dynamics in the SG keeps. Therefore, the stage-3 DAEs are (\ref{eq_stage3DE}) and (\ref{eq_stage3AE}):
\begin{small}	
					\begin{equation}
					\label{eq_stage3DE}
						\left\{ \begin{array}{l}
	\dot{\varphi}_{pg}=\omega _0\omega _{pg}\\
	\dot{\omega}_{pg}=\left( k_{ppll}\dot{u}_{wtq}+k_{ipll}u_{wtq} \right) /\omega _0 - \dot{\omega}_{sg}\\
	\dot{\omega}_{sg}=\left[ P_{msg}-P_{sg}-D_{sg}\left( \omega _{sg}-1 \right) \right] /\left( 2H_{sg} \right)\\
	\dot{i}_{rd}=K_{ramp}\\
						\end{array} \right.
					\end{equation}	
				\end{small}
			and
			\begin{small}
					\begin{equation}
					\label{eq_stage3AE}
						\left\{ \begin{array}{l}
	i_{rq}=i_{rq,norm}\\
	I_r=\sqrt{{i_{rd}}^2+{i_{rq}}^2}\\
	\varphi _I=\arctan \left( i_{rq}/i_{rd} \right)\\
	u_{wtq}= -AE_{sg}\sin \left( \varphi_{pg}-\phi _a \right) + BI_r\sin \left( \phi _b+\varphi _I \right)  \\
	P_{sg}={CE_{sg}}^2\cos \phi _c-E_{sg}DI_r\cos \left[ \varphi _{pg}+\left( \phi _d+\varphi _I \right) \right]\\
						\end{array} \right.
					\end{equation}
				\end{small}
	where $K_{ramp}$ is the ramping rate of active current of the DFIG.	Now the stage-3 dynamics can be viewed as a series of quasi-steady-state with the same form of stage-2 dynamics, but under different values of ${i}_{rd}$ (i.e., ${i}_{rd}$ increases slowly from ${i}_{rd2}$ to ${i}_{rd4} = {i}_{rd1}$) and also different network parameters.
	

\subsection{Simulation verification} 

\begin{figure}[t]
				\centering
						\includegraphics[width=0.9\linewidth]{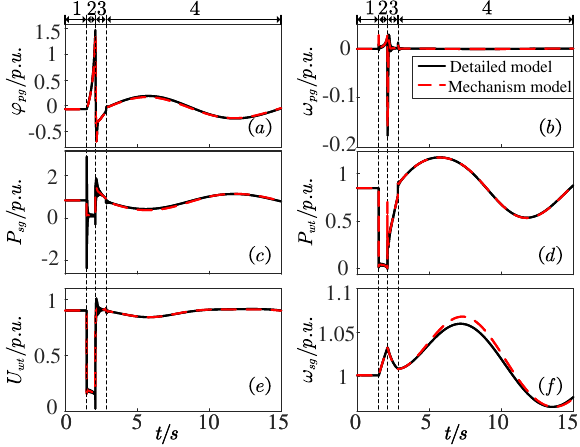}
						\vspace{-0.2cm}
				\caption{(a)-(f) Plots of $\varphi_{pg}$, $\omega_{pg}$, $P_{\rm\mathit{sg}}$, $P_{\rm\mathit{wt}}$, $U_{wt}$, and $\omega_{sg}$, respectively, for comparison of the 4-stage mechanism model and the detailed EMT simulation.}
				\label{Compare_Sim}
				\vspace{-0.3cm}
			\end{figure}

	 The time-domain simulation results of the above 4-stage mechanism model and the detailed EMT model in MATLAB/Simulink are compared in \figurename \ref{Compare_Sim}. The detailed EMT model considers the fast dynamics such as AC current control dynamics, line dynamics, and flux linkage dynamics. At $t_f$ = 1.5 s, a ground fault occurs at the common point, $Z_{L2}$ = 0.01 p.u. for the during-fault. ($Z_{L1}$ = $Z_{L3}$ = $Z_{L4}$ = 0.5 p.u..) At $t_c$ = 2.1 s, the fault is cleared. The fault duration time is $t_c$ – $t_f$ = 0.6 s. During the fault, ${i\rm_{\mathit{rq}2}}=-1.0$ p.u. is calculated based on (\ref{eq_stage2AE}), and ${i\rm_{\mathit{rd}2}}=0.3$ p.u. is chosen. These results clearly indicate that the mechanism model is consistent with the detailed model, although some discernible fast dynamics at the switching moments of each stage are missed.

\section{GSE-based model}

 		\subsection{Entering LVRT condition}
 				
%

\begin{figure}[t]
 					\centering
 							\includegraphics[width=0.7\linewidth]{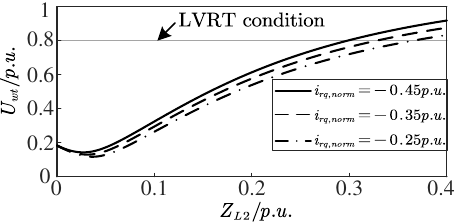}
 									\vspace{-0.2cm}
 					\caption{Plot of terminal voltage $U_{wt}$ at the initial moment of stage 2 under a ground fault, to show that the system always goes into the LVRT easily.}
 					\label{Condition}
 					\vspace{-0.2cm}
 				\end{figure}

	As the first step, we should examine the entering LVRT condition under $U_{wt}< $  0.8 p.u., based on the grid code. We calculate $U_{wt}$ ($U_{wt}=\sqrt{{u_{wtd}}^2+{u_{wtq}}^2}$) of the parallel system at the initial moment of stage 2 under a ground fault. The results are shown in \figurename \ref{Condition}, where different initial reactive current $i_{rq,norm}$ and three-phase grounding resistance $Z_{L2}$ at stage 2 affect the terminal voltage $U_{wt}$. It can be seen that the less $i_{rq,norm}$, the easier it is to enter the LVRT. However, as the resistance of a three-phase grounding fault is usually less than 0.1 p.u., the system should always go into the LVRT. Other network parameters have also been widely studied, and this result is unchanged. Therefore, the LVRT dynamics should always be considered, while it has been ignored in nearly all existing relevant studies \cite{Ref_hybrid1,Ref_hybrid2,Ref_Qu, Ref_TW,Ref_hybrid3}.
	  
	\subsection{Generalized swing equation}
	
	 	\begin{figure}[!t]
		\centering
		\includegraphics[width=0.7 \linewidth]{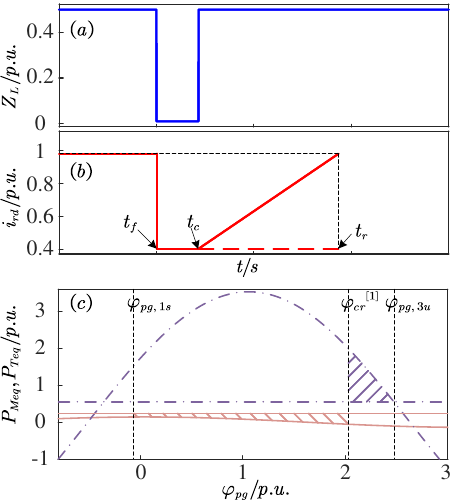}
		\vspace{-0.2cm}
		\caption{
			(a) and (b) Parameter changes of $Z_L$ and ${i\rm_{\mathit{rd}}}$ in the SG-DFIG parallel system. The ${i\rm_{\mathit{rd}}}$ at stage 3 can be virtually viewed as staying at ${i\rm_{\mathit{rd}2}}$ (indicated by a dashed horizontal line).
			In particular, based on the fact that the TSS stability is determined at the beginning time of stage 3, the ${i\rm_{\mathit{rd}}}$ at stage 3 can be virtually viewed as staying at ${i\rm_{\mathit{rd}2}}$ (indicated by a dashed horizontal line) for ever.
			(c) The classical EAC analysis on the GSE. For details, see the text.}					
		\label{EAC}
	\end{figure}


Let us establish the GSE for the SG-DFIG parallel system. According to (\ref{eq_stage2DE})-(\ref{eq_stage2AE}) for the stage-2 dynamics, after inputting $\dot{\omega}_{sg}$ into the equation of $\dot{\omega}_{pg}$, we obtain the pure differential equation, 
				\begin{equation}
				\label{eq_GSE2th}
					\left\{ \begin{array}{l}
	\dot{\varphi}_{pg}=\omega _0\omega _{pg}\\
	\dot{\omega}_{pg}=P_{Meq}-P_{Teq}-D_{eq}\dot{\varphi}_{pg}\\
					\end{array} \right.
				\end{equation}
where $P_{Meq}$ and $P_{Teq}$ are the equivalent mechanical power and the equivalent electromagnetic power, respectively, and $D_{eq}$ is the equivalent nonlinear damping. Their detailed expressions are
\begin{small}
				\begin{equation}
				\label{eq_coe2th}
					\left\{				 \begin{aligned}
					P_{Meq}&=\frac{k_{ipll}}{\omega _0}BI_r\sin \left( \phi _b+\varphi _I \right)-({P_{msg}-{CE_{sg}}^2\cos \phi _c})/({2H_{sg}})\\
					P_{Teq}&=\frac{k_{ipll}}{\omega _0}AE_{sg}\sin \left( \varphi _{pg}-\phi _a \right) \\
					& \quad +E_{sg}DI_r\cos \left( \varphi _{pg}+\phi _d+\varphi _I \right) /\left( 2H_{sg} \right)\\
					&=K_{eq}\sin \left( \varphi _{pg}+\gamma  \right) \\
					D_{eq}&=\frac{k_{ppll}}{\omega _0}AE_{sg}\cos \left( \varphi _{pg}-\phi _a \right)\\
									\end{aligned} \right.
				\end{equation}	
			\end{small}					
with $K_{eq}$, $\gamma$, $m$, and $n$ in $P_{Teq}$ all constant coefficients:	
\begin{small}	
				\begin{equation}
				\label{eq_coePteq}            
									\left\{ \begin{array}{l}
					K_{eq}=\sqrt{{m}^2+{n}^2}\\[2mm]
					\gamma =\arctan \left( n/m \right)\\[2mm]
					m=\dfrac{k_{ipll}}{\omega _0}AE_{sg}\cos \phi _a-E_{sg}DI_r\sin \left( \phi _d+\varphi _I \right) /\left( 2H_{sg} \right)\\[2mm]
					n=E_{sg}DI_r \cos \left( \phi _d+\varphi _I \right)/\left( 2H_{sg} \right)-\dfrac{k_{ipll}}{\omega _0}AE_{sg}\sin \phi _a\\
									\end{array} \right.
				\end{equation}	
\end{small}				
Here due to the relatively small changes of $\omega_{sg}$ in stage 2, as shown in \figurename \ref{Compare_Sim}(f), 
the SG damping term in (\ref{eq_coe2th}) has been neglected, the same as in the traditional power system analysis \cite{Ref_kundur}.

Ultimately, (\ref{eq_GSE2th}) can be expressed as
				\begin{equation}
				\label{eq_GSEHy}
\frac{\ddot{\varphi}_{pg}}{\omega _0}=P_{Meq}-K_{eq}\sin \left( \varphi _{pg}+\gamma  \right) -D_{eq}\dot{\varphi}_{pg}
				\end{equation}

According to (\ref{eq_stage3DE})-(\ref{eq_stage3AE}) for the stage-3 dynamics, based on the quasi-steady-state and local dynamical characteristics near the controlling unstable equilibrium point, the dominant stage-3 dynamics is the same as stage-2 dynamics, and the system fate of TSS is completely determined at the beginning time of stage 3. In the TSS analysis, it is reasonable to treat the whole stage-3 dynamics under the same initial internal parameters of stage 3. In contrast, the external parameters $Z_L$ under the switch change. See Refs. \cite{Ref_ZYY2,Ref_xhs,Ref_Shen_Qihao,Ref_han} for more details. To visualize this key effect, Figs. \ref{EAC}(a) and (b) show the parameter changes of $Z_L$ (as an external parameter) and ${i\rm_{\mathit{rd}}}$ (as an internal parameter) in the LVRT. In particular, the ${i\rm_{\mathit{rd}}}$ at stage 3 has been virtually plotted as staying at ${i\rm_{\mathit{rd}2}}$ for ever (indicated by a dashed horizontal line), not the original linear increase (indicated by a solid sloped line). After this treatment, the stage-3 dynamics could also be described by the same GSE. After the beginning time of stage 3, all the following dynamical behaviors of stages 3 and 4 are unimportant in the TSS analysis. 

On the other hand, for the stage-1 dynamics, although it is a high-dimensional nonlinear system in (\ref{eq_stage1DE})-(\ref{eq_stage1AE}), it only provides an initial (stable) equilibrium point, and hence the GSE in (\ref{eq_GSEHy}) is workable in the analysis. As a result, in the TSS analysis, the whole dynamical process of the parallel system can be described by the same GSE under the different parameter changes in a unified manner, which is schematically shown in Figs. \ref{EAC}(a) and (b). This is similar to what we have done for the single REPG system \cite{Ref_ZYY2,Ref_xhs,Ref_Shen_Qihao,Ref_han}, and it is also similar to the traditional power system. Next, we will mainly use the unified GSE model in the TSS analysis.

\section{TSS analysis and Assessment}
\subsection{Classical EAC method}			
					
			Obviously, the second-order GSE (\ref{eq_GSEHy}) is similar to the SE. Although its coefficients are very complicated, they are all constants, based on the values of $Z_L$, ${i\rm_{\mathit{rq}}}$, and ${i\rm_{\mathit{rd}}}$ at three different stages 1, 2, and 3. As the first approximation, by removing the damping term [i.e., $D_{eq}=0$ in (\ref{eq_GSEHy})], it is ready to use the classical EAC method in the TSS analysis.

			As shown in \figurename \ref{EAC}(c), the solid horizontal line and the dash-dotted horizontal line represent the equivalent mechanical power ($P_{{Meq2}}$ and $P_{{Meq3}}$) of stages 2 and 3, respectively. The  solid sine curve and the dash-dotted sine curve represent the equivalent electromagnetic power ($P_{Teq2}$ and $P_{Teq3}$) of stage 2 and stage 3, respectively. In addition, at stage 1 and the initial moment of stage 3, according to the second equation in (\ref{eq_uwtdq}), $u_{wtq}$ = 0 for the equilibrium point. Thus, the stable equilibrium point of stage 1 ($\varphi_{pg,1s}$) and the unstable equilibrium point of the virtual stage 3 ($\varphi_{pg,3u}$) can be obtained, i.e., 
			\begin{small}
						\begin{equation}
							\label{eq_Operation}
							\left\{ \begin{array}{l}
	\varphi _{pg,1s}=\arcsin \left[{B_1I_{r1}\sin \left( \phi _{b1}+\varphi _{I1} \right)}/{(A_1E_{sg})} \right] +\phi _{a1}\\[4mm]
	\varphi _{pg,3u}=\pi -\arcsin \left[ {B_3I_{r3}\sin \left( \phi _{b3}+\varphi _{I3} \right)}/{(A_3E_{sg})} \right] -\phi _{a3}\\
							\end{array} \right.
						\end{equation}
					\end{small}
Clearly, $\varphi _{pg,1s}$ provides the initial condition and $\varphi _{pg,3u}$ provides the rightest restriction for the worst case, as shown in \figurename \ref{EAC}(c). Here, the network parameters and rotor currents are different at different stages, and they are distinguished by subscript numbers explicitly.

			In \figurename \ref{EAC}(c), based on the EAC, the accelerating area ${S\rm_{\mathit{acc}2}}$ in stage 2 (starting at $\varphi_{pg,1s}$ and ending at the critical clearing angle ${\varphi _{cr}}^{\left[ 1 \right]}$, indicated by the left red area) and the decelerating area ${S\rm_{\mathit{dec}3}}$ in stage 3 (starting at ${\varphi _{cr}}^{\left[ 1 \right]}$ and ending at $\varphi_{pg,3u}$, indicated by the right purple area) are, respectively,  
				\begin{equation}
				\label{eq_EAC1}
				\left\{ \begin{array}{l}
{S_{acc 2}}=\int_{\varphi _{pg,1s}}^{{\varphi _{cr}}^{\left[ 1 \right]}}{[P_{Meq2}-K_{eq2}\sin\left( \varphi _{pg}+\gamma _2 \right)] d\varphi _{pg}}
\\ [3mm]
{S_{dec 3}}=\int_{{\varphi _{cr}}^{\left[ 1 \right]}}^{\varphi _{pg,3u}}{[K_{eq3}\sin\left( \varphi _{pg}+\gamma _3 \right)-P_{Meq3}] d\varphi _{pg}}
				\end{array} \right.
				\end{equation}		
 						
 			For the critical stability, ${S\rm_{\mathit{acc}2}} = {S\rm_{\mathit{dec}3}}$. Correspondingly, the CCA, ${\varphi _{cr}}^{\left[ 1 \right]}$, can be analytically calculated based on (\ref{eq_EAC1}). Next combining the numerical calculation of during-fault trajectory, the CCT can be obtained. The results of the CCA and CCT based on the EAC method are presented in Tables I and II, respectively. By comparing with the EMT numerical simulation results, it can be seen that the error of the EAC method is relatively large (with the CCA error around -14\%, and the CCT error around -6\%), and the more severe the fault 
 			(for a smaller $Z_{\rm_{\mathit{L}2}}$ and a larger $i_{\rm_{\mathit{rd}2}}$), the greater the error.

\subsection{Improved EAC method}					
				Due to the participation of the PLL, the parallel system exhibits the \textquotedblleft frequency jump\textquotedblright \hspace{0.2em}and \textquotedblleft nonlinear damping\textquotedblright effects. Next, we consider these two effects and propose an improved EAC method, which is completely analytically computable.
				
		\textbf{Step 1: Considering frequency jump}

		Firstly, regarding the GSE (\ref{eq_GSEHy}), an energy function is constructed by ignoring the nonlinear damping again ($D_{eq}=0$), i.e.,
														 \begin{equation}
														 		\label{eq_Hamilton}
H=\frac{1}{2}\omega _0{\omega _{pg}} ^2-P_{Meq}\varphi_{pg} -K_{eq}\cos \left( \varphi_{pg} +\gamma \right) 
														 \end{equation}
where the first term denotes the kinetic energy, the second one denotes the position potential, and the third one denotes the electromagnetic potential \cite{Ref_kundur}. 
												 	
		At the moment of fault occurrence (switching from stage 1 to 2) and clearance (switching from stage 2 to 3), the frequency of the SG ($ \omega _{sg}$) does not jump, while the frequency of the PLL ($ \omega _{pll}$) does. Therefore, the change in the frequency mismatch ($\varDelta \omega _{pg}$) at each switch is		
								 \begin{equation}
														 		\label{eq_wpg}
\varDelta \omega _{pg}=\varDelta \omega _{pll}=k_{ppll}/\omega _0\varDelta u_{wtq}
								 \end{equation}

		The frequency jump variables at the initial moments of stages 2 and 3 are denoted as $\varDelta \omega_{pg2}$ and $\varDelta \omega_{pg3}$, respectively. The superscripts ($+$) and ($-$) indicate the moments before and after a certain time; for example, ${u_{wtq2}}^-$ (${u_{wtq2}}^+$) represents the $q$-axis terminal voltage of the DFIG before (after) the initial moment of the stage 2. At the initial moment of stage 2, the $\varphi _{pg1,s}$ does not change, while the network parameters and internal parameters do. The expression of $\varDelta \omega _{pg2}$ is		
														 \begin{equation}
														 		\label{eq_dwpg2}
\varDelta \omega _{pg2}=k_{ppll}/\omega _0\left( {u_{wtq2}}^+-{u_{wtq2}}^- \right)
														 \end{equation}			
		and, according to (\ref{eq_uwtdq}), ${u_{wtq2}}^-$ and ${u_{wtq2}}^+$ are								
											\begin{equation}
											\label{eq_uwtq2}
												\left\{				 \begin{aligned}
												{u_{wtq2}}^-=&0\\ 
												{u_{wtq2}}^+=&B_2I_{r2}\sin ( \phi _{b2}+\varphi _{I2} )-A_2E_{sg}\sin ( \varphi _{pg,1s}-\phi _{a2} )\\
																\end{aligned} \right.
											\end{equation}
	
		Similarly, the expression of $\varDelta \omega _{pg3}$ is												
															 \begin{equation}
															 		\label{eq_dwpg3}
	\varDelta \omega _{pg3}=k_{ppll}/\omega _0\left( {u_{wtq3}}^+-{u_{wtq3}}^- \right)
															 \end{equation}	
				and ${u_{wtq3}}^-$ and ${u_{wtq3}}^+$ are	
												\begin{equation}
												\label{eq_uwtq3}
								\left\{				 \begin{aligned}
								{u_{wtq3}}^-=&B_2I_{r2}\sin \left( \phi _{b2}+\varphi _{I2} \right) -A_2E_{sg}\sin (  {\varphi _{cr}}^{\left[ 2 \right]} -\phi _{a2} ) \\ 
								{u_{wtq3}}^+=&B_3I_{r3}\sin \left( \phi _{b3}+\varphi _{I3} \right)-A_3E_{sg}\sin (  {\varphi _{cr}}^{\left[ 2 \right]} -\phi _{a3} )  \\
																	\end{aligned} \right.
												\end{equation}
where ${\varphi _{cr}}^{\left[ 2 \right]}$ denotes the new CCA prediction.

	The initial energy $h_2$ of stage 2 is influenced by the \textquotedblleft frequency jump\textquotedblright, and based on (\ref{eq_Hamilton}), the expression of $h_2$ is
								\begin{equation}
															 		\label{eq_h2}
h_2=\frac{1}{2}\omega _0{\varDelta \omega _{pg2}}^2-P_{Meq2}\varphi _{pg,1s}-K_{eq2}\cos \left( \varphi _{pg,1s}+\gamma _2 \right) 
								\end{equation}	
		Similarly, the energy $h_3$ at stage 3 can also be obtained, based on the unstable equilibrium point $\varphi_{pg,3u}$:
																	 \begin{equation}
																	 		\label{eq_h3}
				h_3=-P_{Meq3}\varphi _{pg,3u}-K_{eq3}\cos \left( \varphi _{pg,3u}+\gamma _3 \right) 
																	 \end{equation}

		Due to $D_{eq}=0$, the system is a Hamiltonian conservative system during stages 2 and 3. The energy in stage 2 and stage 3 is equal to $h_2$ and $h_3$, respectively. According to (\ref{eq_Hamilton}), (\ref{eq_h2}) and (\ref{eq_h3}), we obtain the frequency mismatch during stages 2 and 3:
									 		\begin{equation}
											\label{eq_dwpg23}
												\left\{				 \begin{aligned}
	{\omega _{pg2}}=\sqrt{{2[ h_2+P_{Meq2}{\varphi _{pg}}+K_{eq2}\cos ( {\varphi _{pg}}+\gamma _2 )]} /{\omega _0}}\\[2mm]
	{\omega _{pg3}}=\sqrt{{2[ h_3+P_{Meq3}{\varphi _{pg}}+K_{eq3}\cos ( {\varphi _{pg}}+\gamma _3 ) ] }/{\omega _0}}\\
																\end{aligned} \right.
											\end{equation}

		Before and after the fault clearance moment, the frequency mismatch are, respectively, represented by ${\omega _{pg3}}^-$ and ${\omega _{pg3}}^+$. Based on (\ref{eq_dwpg23}), the expressions of ${\omega _{pg3}}^-$ and ${\omega _{pg3}}^+$ are

									 		\begin{equation}
											\label{eq_dwpg3b}
												\left\{				 \begin{aligned}
	{\omega _{pg3}}^-=	{\omega _{pg2}}({\varphi _{cr}}^{[2]})\\[2mm]
	{\omega _{pg3}}^+={\omega _{pg3}}({\varphi _{cr}}^{[2]})\\
																\end{aligned} \right.
											\end{equation}

       Here ${\omega _{pg3}}^+$ and ${\omega _{pg3}}^-$ are connected by $\varDelta \omega _{pg3}$, i.e.,
										\begin{equation}
																	 		\label{eq_EAC2}
			{\omega _{pg3}}^+ -{\omega _{pg3}}^- =\varDelta \omega _{pg3}
										\end{equation}
							where both ${\omega _{pg3}}^+$ and ${\omega _{pg3}}^-$ are a function of ${\varphi_{cr}}^{[2]}$, as in (\ref{eq_dwpg3b}), and $\varDelta \omega _{pg3}$ is also a function of ${\varphi_{cr}}^{[2]}$, as in (\ref{eq_dwpg3}) and (\ref{eq_uwtq3}). Therefore, the value of ${\varphi_{cr}}^{[2]}$ can be solved.

		\textbf{Step 2: Considering nonlinear damping}	 
				
				Next, on the basis of Step 1, we will take into account the dissipative energy due to the nonlinear damping term ($D_{eq} \neq 0$). We use the already-obtained result of Step 1 to help estimate. Hence the dissipative energy $S_d$ is expressed as
				\begin{small}
					\begin{equation}
																	 		\label{eq_Sd}
S_d\approx \int_{\varphi _{pg,1s}}^{\varphi _{cr}^{\left[ 2 \right]}}{D_{eq2}\omega _0\omega _{pg2}d\varphi _{pg}}+\int_{\varphi _{cr}^{\left[ 2 \right]}}^{\varphi _{pg,3u}}{D_{eq3}\omega _0\omega _{pg3}d\varphi _{pg}}
					\end{equation}	
				\end{small} 
					where $\omega _{pg2}$ and $\omega _{pg3}$ still use (\ref{eq_dwpg23}) for the approximation.

						\begin{table}[t]
											\begin{center}
												\label{t1}
												\caption{Comparison of CCA under different values of
												$Z_{\rm_{\mathit{L}2}}$ and $i_{\rm_{\mathit{rd}2}}$}
												\scalebox{.96}{
												\setlength{\tabcolsep}{2.56mm}{
												\begin{tabular}{ccccccc}
													\toprule
													\multirow{2}{*}[-1ex]{{\makecell{$Z_{\rm_{\mathit{L}2}}$/$i_{\rm_{\mathit{rq}2}}$\\(p.u.)}}} & \multirow{2}{*}[-1ex]{{\makecell{{$i_{\rm_{\mathit{rd}2}}$}\\(p.u.)}}}& \textbf{EMT} & \multicolumn{2}{c}{\textbf{EAC}} & \multicolumn{2}{c}{\textbf{Improved EAC}}\\
													\cline{3-7}
													& & {{\makecell{{CCA}\\(rad)}}} &{{\makecell{{CCA}\\(rad)}}} &  {\makecell{relative\\error}}  &  {{\makecell{{CCA}\\(rad)}}}&  \makecell{relative\\error} \\  
													\midrule
													\multirow{3}{*}{0.01/-1}    &  0.4 & 2.236  & 2.030 & - 8.1\%  & 2.180 & - 2.5\% \\  
																		   	   &  0.5 & 2.154  & 1.897 & - 11.9\%  & 2.089 & - 3.0\%\\  
																		   	   &  0.6 & 2.074  & 1.777 & - 14.3\%  & 2.009 & - 3.1\%\\  
													\cline{1-7}
													\multirow{3}{*}{0.001/-0.97} &  0.4 & 2.210 & 1.931 & - 12.6\%  & 2.187 & - 1.1\% \\  
																			   &  0.5 & 2.131  & 1.809 & - 15.1\%  & 2.109 & - 1.0\%\\  
																			   &  0.6 & 2.053  & 1.696 & - 17.4\%  & 2.037 & - 0.8\%\\  
																			   
													\bottomrule
												\end{tabular}
												}
												}
											\end{center}
						\end{table}

				\begin{table}[t]
									\begin{center}
										\label{t2}
										\caption{Comparison of CCT under different values of
										$Z_{\rm_{\mathit{L}2}}$ and $i_{\rm_{\mathit{rd}2}}$}
										\scalebox{.96}{
										\setlength{\tabcolsep}{2.56mm}{
										\begin{tabular}{ccccccc}
											\toprule
											\multirow{2}{*}[-1ex]{{\makecell{$Z_{\rm_{\mathit{L}2}}$/$i_{\rm_{\mathit{rq}2}}$\\(p.u.)}}} & \multirow{2}{*}[-1ex]{{\makecell{{$i_{\rm_{\mathit{rd}2}}$}\\(p.u.)}}}& \textbf{EMT} & \multicolumn{2}{c}{\textbf{EAC}} & \multicolumn{2}{c}{\textbf{Improved EAC}}\\
											\cline{3-7}
											& & {{\makecell{{CCT}\\(s)}}}  &{{\makecell{{CCT}\\(s)}}}  &  {\makecell{relative\\error}}  &  {{\makecell{{CCT}\\(s)}}} &  \makecell{relative\\error} \\  
											\midrule
											\multirow{3}{*}{0.01/-1}    &  0.4 & 0.301  & 0.289 & - 4.0\%  & 0.298 & - 1.0\% \\  
																   	   &  0.5 & 0.218  & 0.205 & - 6.0\%  & 0.215 & - 1.4\%\\  
																   	   &  0.6 & 0.175  & 0.162 & - 7.4\%  & 0.172 & - 1.7\%\\  
											\cline{1-7}
											\multirow{3}{*}{0.001/-0.97} &  0.4 & 0.195 & 0.185 & - 5.1\%  & 0.194 & - 1.0\% \\  
																	   &  0.5 & 0.162  & 0.151 & - 6.8\%  & 0.161 & - 1.2\%\\  
																	   &  0.6 & 0.139  & 0.127 & - 8.6\%  & 0.138 & - 1.4\%\\  
																	   
											\bottomrule
										\end{tabular}
										}
										}
									\end{center}
				\end{table}

		Relying on the energy conversation rule and the GSE in (\ref{eq_GSEHy}) and considering the effects of both \textquotedblleft frequency jump\textquotedblright \hspace{0.2em}and \textquotedblleft nonlinear damping\textquotedblright \hspace{0.2em} yield the implicit expression of ${\varphi _{cr}}^{\left[ 3 \right]}$ for the improved CCA,	
		\begin{small}
				\begin{equation}
				\label{eq_EAC3}
				\begin{aligned}
				&\frac{1}{2}\omega _0\left( {\omega _{pg3}}^- \right) ^2
												-\frac{1}{2}\omega _0\left( {\omega _{pg3}}^+ \right) ^2
												 -\frac{1}{2}\omega _0\left( \varDelta \omega _{pg2} \right) ^2 \\	
												=&\int_{\varphi _{pg,1s}}^{\varphi _{cr}^{\left[ 3 \right]}}{[P_{Meq2}-K_{eq2}\sin \left( \varphi _{pg}+\gamma _2 \right)] d\varphi _{pg}}+
												\\
											&\int_{\varphi _{cr}^{\left[ 3 \right]}}^{\varphi _{pg,3u}}{[P_{Meq3}-K_{eq3}\sin \left( \varphi _{pg}+\gamma _3 \right)] d\varphi _{pg}}-S_d
																\end{aligned}
											\end{equation}
	\end{small}

		Based on the result of Step 1 (${\varphi _{cr}}^{\left[ 2 \right]}$), all kinetic energy terms can be calculated according to (\ref{eq_dwpg2}), (\ref{eq_uwtq2}), and (\ref{eq_dwpg3b}). The effect of the damping term $S_d$ is estimated approximately using (\ref{eq_Sd}). Clearly (\ref{eq_EAC3}) is a nonlinear algebraic equation with respect to ${\varphi _{cr}}^{\left[ 3 \right]}$, which can be quickly solved using the Newton-Raphson method.
		The	${\varphi _{cr}}^{\left[ 3 \right]}$ is the ultimate outcome of the improved EAC-based method.

		In Tables I and II, the numerical results and theoretical results of the classical EAC-based method and the improved EAC-based method are compared. Their relative errors with the EMT simulation are compared. 
		It can be seen that as the fault severity increases the active power output of the DFIG during LVRT increases, their CCA and CCT gradually decrease. In the estimation of CCA, the improved EAC method always shows a tiny error within -3\%. While in the CCT estimation, it is always within -2\% error. Both are conservative. Comparatively, the improved EAC results are excellent.

\section{Experimental Verification}	
\begin{figure}[t!]
	\centering
\vspace{-0.2cm}
	\includegraphics[width=0.6 \linewidth]{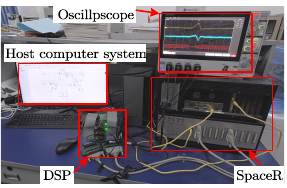}
	\caption{SpaceR real-time simulation platform.}
	\vspace{-0.2cm}
	\label{HIL}
\end{figure}

To verify the correctness of the above theory, hardware-in-the-loop experiments are conducted based on SpaceR. As shown in \figurename \ref{HIL}, the SpaceR real-time simulation platform consists of the host computer system, digital signal processing, oscilloscope and SpaceR. This platform integrates functions such as real-time simulation, control, testing, rapid control prototyping, and hardware-in-the-loop testing. It can compile the target model and then simulate its dynamic behavior. The integration and operation of the object model and the physical control system are achieved through the digital signal processing device on the simulation platform, which facilitates the testing of the physical control system. The control parameters tested in the experiment are the same as in Appendix A. Here, three sets of experiments are presented to verify the correctness of the above mechanism analysis and stability assessment.

\begin{figure}[t!]
	\centering
\vspace{-0.2cm}
	\includegraphics[width=0.6 \linewidth]{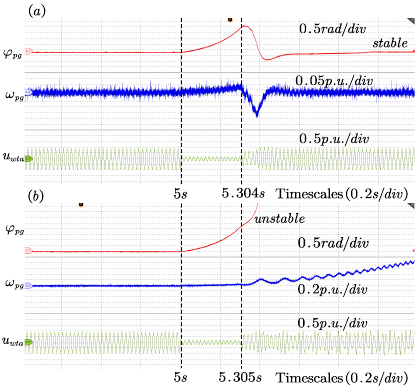}
	\caption{(a) and (b) Experimental waveform diagrams of Case A.}
	\vspace{-0.4cm}
	\label{Case A}
	
\end{figure}

Case A: When $t_f$ = 5 s, $Z_{L2}$ is 0.01 p.u. during fault, $i_{\rm\mathit{rd}2}$ = 0.4 p.u. and $i_{\rm\mathit{rq}2}$ = -1 p.u. At $t_c$, the fault is cleared. The experimental results for two different fault clearing times $t_c$ = 5.304 s and 5.305 s are shown in Figs. \ref{Case A}(a) and (b), respectively. It is clear that with a slightly larger $t_c$ in \figurename \ref{Case A}, although $\varphi_{\rm\mathit{pg}}$ is unstable in stage 2, the system can finally become stable. Under this situation, the CCT is 0.304 s, well in accord with the EMT result: CCT = 0.301 s in Table II. These experimental results demonstrate that even if the system experiences transient instability in stage 2, as long as the fault is cleared in time, the system can ultimately be stable.			

\begin{figure}[t!]
	\centering
	\includegraphics[width=0.6 \linewidth]{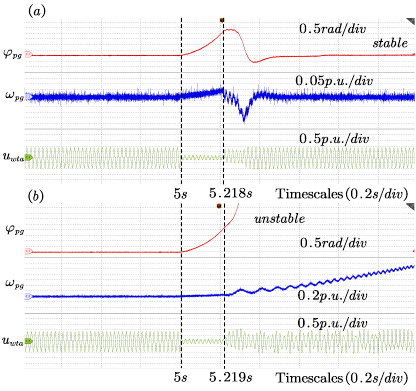}
	\vspace{-0.2cm}
	\caption{(a) and (b) Experimental waveform diagrams of Case B.}
	\label{Case B}
	\vspace{-0.2cm}
\end{figure}

\begin{figure}[t!]
	\centering
	\includegraphics[width=0.6 \linewidth]{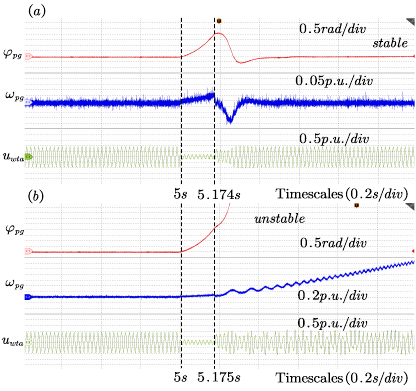}
	\vspace{-0.2cm}
	\caption{(a) and (b) Experimental waveform diagrams of Case C.}
	\label{Case C}
	\vspace{-0.3cm}
\end{figure}
Case B: When $t_f$ = 5 s, $Z_{L2}$ is 0.01 p.u. during fault, $i_{\rm\mathit{rd}2}$ = 0.5 p.u. and $i_{\rm\mathit{rq}2}$ = -1 p.u. At $t_c$, the fault is cleared. The experimental results for two different fault clearing times $t_c$ = 5.218 s and 5.219 s are shown in Figs. \ref{Case B}(a) and (b), respectively. Under this situation, the CCT is 0.218 s, well in accord with the EMT result: CCT = 0.218 s in Table II.

Case C: When $t_f$ = 5 s, $Z_{L2}$ is 0.01 p.u. during fault, $i_{\rm\mathit{rd}2}$ = 0.6 p.u. and $i_{\rm\mathit{rq}2}$ = -1 p.u. At $t_c$, the fault is cleared. The experimental results for two different fault clearing times $t_c$ = 5.174 s and 5.175 s are shown in Figs. \ref{Case B}(a) and (b), respectively. Under this situation, the CCT is 0.174 s, well in accord with the EMT result: CCT = 0.175 s in Table II. Therefore, Cases A, B, and C all well verify the effectiveness of the aforementioned TSS assessment method and prove the correctness of the stability mechanism analysis.

\section{Conclusion}
				In conclusion, for the first time, the TSS model, analysis, and assessment of the SG-DFIG parallel system considering complete LVRT processes have been systematically studied. The necessity and importance to study the LVRT have been well addressed. The assessment results relying on the sequential improved EAC method and the GSE model of three stages 1, 2, and 3 have been found to be perfect. 
				It is lucky to see that in the SG and REPG parallel system, as the simplest hybrid systems, the frequency and phase mismatches of the two devices are two independent dominant variables, and thus, the unified theory including the unified GSE model and the improved EAC method is still workable. This is quite similar to the equivalent single-machine model of the SG-SG parallel system and the EAC analysis.

For discussions, first compared with some other relevant TSS analysis of hybrid systems, where the renewable energy dynamics is usually ignored or algebraic by using the singular perturbation technique \cite{Ref_Tian,Ref_Shen, Ref_GeX, Ref_ChenL1}, here it is found that the REPG after the LVRT shows the electromechanical time-scale dynamics which is apparently comparable with that of SG, and our analyses precisely consider these effects. Second, compared with most of recent studies on the during-fault stability for either existence of equilibrium point or transient stability \cite{Ref_12,Ref_19} and those ignoring the LVRT processes \cite{Ref_hybrid1,Ref_hybrid2,Ref_Qu, Ref_TW,Ref_hybrid3}, the complete LVRT processes have been sufficiently and efficiently considered. Third, it is found that if the role of GSC is considered, the above method is still workable. Last but not the least, different from some recent improved EAC methods, which highly rely on iteration algorithms from trajectory data \cite{Li_Xilin1,Li_Xilin2}, our method is analytical and predictable, by keeping the essence of the EAC: the law of conservation of energy.

				\appendix
				\setcounter{equation}{0}
				\subsection{Parameters used in the paper}
				
				Parameters of grid: $S_{\rm\mathit{base}}$ = 2 MW, $U_{\rm\mathit{base}}$ = 690 V (line rms value), $U_{\rm\mathit{dcbase}}$ = 1400 V, $f_0$ = 50 Hz, $\omega_0$ = 2${\pi}f_0$,   ${\omega_{\rm\mathit{r}}}^{*}$ = 1.2 p.u., $Z_{sg}$ = 0.1$j$ p.u., $Z_{wt}$ = 0.2$j$ p.u., $Z_L$ = 0.5 p.u. 
				
				Parameters of SG: $P\rm_{\mathit{msg}}$ = 0.85 p.u., $E_{\rm\mathit{sg}}$ = 1.01 p.u., $H_{\rm\mathit{sg}}$ = 6 p.u., $D_{\rm\mathit{sg}}$ = 5 p.u.
				
				Parameters of DFIG: $P\rm_{\mathit{mwt}}$ = 0.85 p.u., $X_{\rm\mathit{s}}$ = 4.071 p.u., $X_{\rm\mathit{r}}$ = 4.056 p.u., $X_m$ = 3.9 p.u., $H_{\rm\mathit{wt}}$ = 6 p.u.

				Parameters of controller: (1) RSC: $k_{\rm\mathit{pw}}$ = 1, $k_{\rm\mathit{iw}}$ = 5. (2) Normal reactive current: $i_{\rm\mathit{rq,norm}}$ = -0.25 p.u.. (3) PLL: $k_{\rm\mathit{ppll}}$ = 60, $k_{\rm\mathit{ipll}}$ = 1400. (4) ACC: $k_{\rm\mathit{purd}}$ = 0.6, $k_{\rm\mathit{iurd}}$ = 80, $k_{\rm\mathit{purq}}$ = 0.6, $k_{\rm\mathit{iurq}}$ = 80. (5) LVRT: $K_e$ = 1. (6) Ramp: $K_{\rm\mathit{ramp}}$ = 0.8.
			
				\subsection{Equivalent single-machine model of SG-SG system} 
\begin{figure}[t!]
						\centering
						\includegraphics[width=0.7\linewidth]{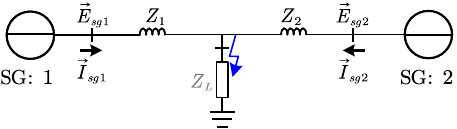}
						\vspace{-0.2cm}
						\caption{Topology diagram of the SG-SG system.}
						\label{SG_SG}
					\end{figure}

				\renewcommand\theequation{B.\arabic{equation}}
					Figure \ref{SG_SG} shows the two SGs connected in parallel to a constant impedance \cite{Ref_kundur}. Usually dynamic aggregation can be utilized to simplify a multi-machine system into the  similar two-cluster system, and this model is very important in the power system analysis \cite{Ref_Ener, Ref_XueY}. The equations described by the second-order rotor swing equation are
				\begin{small}
							\begin{equation}
								\label{eq_2SG}
								\left\{ \begin{array}{l}
	\dot{\theta}_{sg1}=\omega _0\omega _{sg1}\\[2mm]
	\dot{\omega}_{sg1}={[P_{msg1}-P_{sg1}-D_{sg1}\left( \omega _{sg1}-1 \right)]}/{(2H_{sg1})}\\[2mm]
	\dot{\theta}_{sg2}=\omega _0\omega _{sg2}\\[2mm]
	\dot{\omega}_{sg2}=[{P_{msg2}-P_{sg2}-D_{sg2}\left( \omega _{sg2}-1 \right)]}/{(2H_{sg2})}\\
								\end{array} \right.
							\end{equation}
				\end{small}
				where $P_{msgi}$, $D_{sgi}$, and $H_{sgi}$ ($i$ = 1, 2) represent the mechanical power, damping, and inertia of the $i$-th machine.
			
			According to the circuit equation, the relationship between the output current and the internal voltage is
			\begin{small}
									\begin{equation}
\left[ \begin{array}{c}
	\vec{I}_{sg1}\\
	\vec{I}_{sg2}\\
\end{array} \right] =\boldsymbol{Y} \left[ \begin{array}{c}
	\vec{E}_{sg1}\\
	\vec{E}_{sg2}\\
\end{array} \right] 
										\label{eq_2SG_cir}
									\end{equation}
								\end{small}
by the admittance matrix
											\begin{small}
												\begin{equation}
\boldsymbol{Y}=\left[ \begin{matrix}
	Y_{11}&		Y_{12}\\
	Y_{21}&		Y_{22}\\
\end{matrix} \right] =\left[ \begin{matrix}
	Z_L+Z_1&		Z_L\\
	Z_L&		Z_L+Z_2\\
\end{matrix} \right] ^{-1}
													\label{eq_2SG_cirmatrix}
												\end{equation}
											\end{small}
			where $Z_1$ and $Z_2$ are line impedances, and $Z_L$ is load impedance. In addition, ${Y_{12}} = {Y_{21}}$.
			
			The electromagnetic powers for the two SGs are 
			\begin{small}
										\begin{equation}
											\label{eq_2SG_Psg}
											\left\{ \begin{array}{l}
	P_{sg1}=\mathrm{Re}\left[ \vec{E}_{sg1} \overline{\left( Y_{11}\vec{E}_{sg1}+Y_{12}\vec{E}_{sg2} \right) } \right]\\[3mm]
	P_{sg2}=\mathrm{Re}\left[ \vec{E}_{sg2} \overline{\left( Y_{12}\vec{E}_{sg1}+Y_{22}\vec{E}_{sg2} \right) } \right]\\
											\end{array} \right.
										\end{equation}
			\end{small}
			where $\vec{E}_{sg1}=E_{sg}\angle \theta _{sg1}$, $\vec{E}_{sg2}=E_{sg}\angle \theta _{sg2}$.	For the admittance, taking $Y_{11}$ as an example, $Y_{11}=\left| Y_{11} \right|\angle \phi _{11}=G_{11}+jB_{11}$, $G_{11}=\left| Y_{11} \right|\cos \phi _{11}$, $B_{11}=\left| Y_{11} \right|\sin \phi _{11}$.
			
			Taking the phase and frequency differences ($\varphi _{12}/\omega _{12}$) as the state variables, 
										\begin{equation}
											\label{eq_2SG_dif}
											\left\{ \begin{array}{l}
				\varphi _{12}=\theta _{sg1}-\theta _{sg2}\\
					\omega _{12}=\omega _{sg1}-\omega _{sg2}\\
											\end{array} \right.
										\end{equation}			
		 Eqs. (\ref{eq_2SG_Psg}) can be further simplified as follows:
										\begin{equation}
											\label{eq_2SG_Psgsim}
											\left\{ \begin{array}{l}
					P_{sg1}={E_{sg}}^2G_{11}+{E_{sg}}^2\left| Y_{12} \right|\cos \left( \varphi _{12} -\phi _{12} \right)\\
					P_{sg2}={E_{sg}}^2G_{22}+{E_{sg}}^2\left| Y_{21} \right|\cos \left( \varphi _{12} + \phi _{21} \right)\\
											\end{array} \right.
										\end{equation}			
			where $G_{11}\ne G_{22}$, $\left| Y_{12} \right|=\left| Y_{21} \right|$, $\phi _{12}=\phi _{21}$.
			
			When the uniform condition ($D_{sg1}/H_{sg1} = D_{sg2}/H_{sg2}= D_{sg}/H_{sg}$) (or the zero damping condition: $D_{sg1}=D_{sg2}=0$ for the first swing study) is met, (\ref{eq_2SG}) can be simplified the same as the single-SG SE model,
			\begin{small}
							\begin{equation}
								\label{eq_2SG_SimDE}
					\left\{ \begin{array}{l}
					\dot{\varphi}_{12}=\omega _0\omega _{12}\\[2mm]
					\dot{\omega}_{12}=\dfrac{P_{msg1}-P_{sg1}}{2H_{sg1}}-\dfrac{P_{msg2}-P_{sg2}}{2H_{sg2}}-\dfrac{D_{sg}}{2H_{sg}}\omega _{12}\\
					\end{array} \right.
													\end{equation}	
			\end{small}
or
\begin{small}
									\begin{equation}
\frac{\ddot{\varphi}_{12}}{\omega _0}=P_{Meq}-P_{Teq}-\dfrac{D_{sg}}{2H_{sg}}\dfrac{\dot{\varphi}_{12}}{\omega _0}
										\label{eq_2SG_SE}
									\end{equation}	
\end{small}			
and further the classical EAC can be applied  \cite{Ref_kundur}.
			
			


				
				%
				\ifCLASSOPTIONcaptionsoff
				\newpage
				\fi

				
				
				%

				\bibliographystyle{IEEEtran}
				\bibliography{ref-V5}

\vfill				
			\end{document}